# mLS-GKM: Efficient Multi-class Regulatory Sequence Classification with Gapped k-mer SVMs

**Kieran Howard, Nathan Harmston**

## Abstract

**Summary:** Gapped k-mer support vector machines (gkm-SVMs) are widely used for classifying regulatory DNA sequences and identifying the sequence features underlying those predictions. Although LS-GKM provides an efficient implementation of gkm-based kernels, it is restricted to binary classification and does not provide calibrated probability outputs. Here, we present mLS-GKM, an extension of LS-GKM that adds multiclass classification, probability-calibrated predictions, parallelised inference, memory efficient sequence interpretation and checkpointing during training. Across 322 ENCODE ChIP-seq datasets, classifiers trained using mLS-GKM are identical to those produced by LS-GKM, while gkmpredict and gkmexplain run 22x and 80x faster respectively at 64 threads, and gkmexplain peak memory is reduced by more than 65%. As a practical demonstration, mLS-GKM was able to accurately distinguish between enhancers, promoters and CTCF binding sites directly from sequence and identified biologically relevant regulatory motifs. Together, these improvements extend gkm-SVMs to multiclass problems and substantially improve their scalability, enabling efficient interpretation of regulatory DNA sequences in large and complex datasets.

**Availability and implementation**: mLS-GKM is implemented in C/C++ and is available at:
https://github.com/harmstonlab/mLS-GKM

**Contact**: harmstonn@cardiff.ac.uk

Gapped k-mer SVMs are used for predicting cis-regulatory activity directly from DNA sequence (Ghandi et al., 2014), including enhancer activity, silencer elements (Doni Jayavelu *et al.*, 2020), and the effects of non-coding variants (Beer, 2017). LS-GKM (Lee, 2016) implemented the gkm kernel inside the LIBSVM decomposition framework (Chang and Lin, 2011), enabling its application to larger datasets and introducing the *gkmrbf*, *wgkm* and *wgkmrbf* kernels. Two companion tools, gkmpredict and gkmexplain (Shrikumar, Prakash and Kundaje, 2019) score new sequences and decompose those scores into per-base contributions, respectively. These tools have been applied in a variety of applications, including prediction of non-coding variant impacts (VandenBosch *et al.*, 2022) and TF-DNA binding (Han *et al.*, 2024).

Despite extensive use, LS-GKM remains limited in three ways that limit its application to larger and more complex problems. First, it supports only binary classification, so distinguishing among more than two classes requires fitting, evaluating and combining the output of multiple binary models. Second, it produces only uncalibrated decision values, which complicates downstream tasks where probabilities are needed (e.g. thresholding or comparison with other classifiers). Third, in the LS-GKM implementation of gkmpredict the choice of threads is limited to 1, 4 or 16 for kernel-internal parallelism, and gkmexplain is single-threaded. Additionally, gkmpredict creates the full per-position-per-base contribution matrix before collapsing it into a final score, which makes interpretation of long sequences or large datasets impractical. In addition, on-disk checkpointing was implemented so that long running training jobs can recover from hardware failures and cluster-imposed wall-clock limits. mLS-GKM was developed to address all these limitations enabling its use on large and complex problems.

## Multi-class training and prediction

mLS-GKM extends gkmtrain and gkmpredict from binary to multiclass (K > 2) classification using LIBSVM's standard one-vs-one comparisons, trained on top of the unmodified LS-GKM kernels. For K classes, *K(K-1)/2* binary sub-problems are solved and stored as a single multi-class model file with a header that records all class labels, per-class support-vector counts, and per-pair bias terms. Prediction proceeds either by returning the pairwise decision boundaries for each comparison or by returning a calibrated probability over all classes.

Calibrated probabilities are obtained by extending LIBSVM's probability machinery to the gkm kernel family. During training, with each of the *K(K-1)/2* pairwise classifiers undergoes an internal 5-fold cross-validation; out-of-fold decision values are pooled and a per-pair sigmoid is fitted by Newton-step Platt scaling with smoothed targets. At prediction time, pairwise decision values are mapped to pairwise conditional probabilities and subsequently combined into a single class-probability vector using the iterative coupling algorithm of Wu, Lin and Weng (Wu, Lin and Weng, 2004). The output of -P therefore has a probabilistic interpretation that summing pairwise margins does not, improving interpretability and providing a measure of confidence for each prediction.

Multi-class extension does not affect models in the binary case: training a two-class problem with mLS-GKM produces a model that is mathematically identical to LS-GKM, the only difference being that internal class labels are stored as 1 and 2 rather than 1 and −1. Across all 322 ENCODE ChIP-seq datasets considered (The ENCODE Project Consortium, 2012), support vectors and their α coefficients were bit-identical between the two implementations (Fig. 1A).

## Parallel prediction and streaming explanation

To improve scalability and performance, both gkmpredict and gkmexplain in mLS-GKM have been parallelised over input sequences using POSIX threads. Each worker thread pulls FASTA records from a shared producer queue, computes its prediction or contribution scores independently using the existing kernel evaluator, and writes its result to a pre-indexed output slot, so that final output order is preserved without locking. Users can specify the number of threads via the new -t flag. For gkmpredict, parallelising over sequences is approximately twice as efficient as the existing multithread kernel computations (Fig. 1B), while also allowing any number of threads to be used with only a marginal increase in memory overhead (Fig. S5).

mLS-GKM also introduces a streaming implementation of gkmexplain. The original implementation builds, for each sequence, an L × 4 matrix of per-position, per-base contributions for each SV, before collapsing this tensor to a per-position per-base score (Shrikumar, Prakash and Kundaje, 2019). For long sequences, large test sets or SVMs with a high number of SVs, this can lead to substantial memory usage. mLS-GKM instead accumulates the contribution of each (position, base) entry directly into the running per-base score as it is computed, avoiding the instantiation of the intermediate tensor. Output is bit-identical to the original gkmexplain but space complexity is reduced from O(L x SV) to O(L + SV), which in practice corresponds to a >65% reduction in memory usage and a 2.8x increase in speed before multithreading is applied (Fig. 1C,D).

## Checkpointing for long-running calibrated jobs

Probability-calibrated training is intrinsically expensive, as each pairwise classifier must be trained six times (once on the full pair, plus five cross-validation folds for the sigmoid fit). For large datasets, training time may exceed typical HPC wall-clock limits. Therefore, mLS-GKM introduces optional incremental checkpointing, allowing interrupted training to resume automatically from the last completed point; specified using the -C argument. Checkpoints are validated against the current configuration before resuming and removed following successful completion.

**Benchmarks**

The performance of mLS-GKM was evaluated on 322 ENCODE ChIP-seq datasets (≥ 5,000 peaks each), using the previously described training/test partitioning strategy (Lee, 2016), in which chromosomes 1 and 2 were held out for testing. Negative sequences were sampled from the genome to match GC-content and repeat fraction. Models were trained using the gkmrbf kernel (-t 3) with recommended hyperparameters (-c 10 -g 2, -m 4000, -T 4) (see Supplementary Methods S1).

**Equivalence in the binary setting**. Across all 322 datasets, ROC AUCs from mLS-GKM gkmpredict were identical to those obtained using LS-GKM gkmpredict (Fig. 1A; Pearson r = 1.00, |ΔAUC| = 0).

**Speed**. On the H1hescCtcf dataset (n = 55,782 peaks, 18,876 test sequences), gkmpredict achieved 22x speed-up using 64 threads compared to single-threaded LS-GKM (Fig. 1B). Scaling for gkmexplain on a large synthetic dataset reached 80x at 64 threads (Fig. 1C). Performance gains plateaued between 64 and 128 threads.

**Memory**. On the same H1hescCtcf dataset, gkmexplain peak resident set size dropped from 1.34 GB (LS-GKM) to 460 MB (mLS-GKM), a 65% reduction in memory usage (Fig. 1D).

**Application: distinguishing regulatory element classes from sequence alone**

To demonstrate the practical applications of mLS-GKM, a single 3-class classifier was trained to distinguish between three common classes of regulatory element (Enhancer, Promoter, CTCF Binding Site) in the human genome (chromosomes 1-2 held out for testing). Performing this analysis with LS-GKM would require training three independent binary classifiers and ad hoc combination of their outputs. In contrast, mLS-GKM performs training, prediction and probability calibration within a single workflow.

The resulting classifier achieved one-vs-rest ROC AUCs of 0.958 (Enhancer), 0.889 (Promoter) and 1.000 (CTCF) on the held-out test set (Fig. 1E). The lower performance for promoters reflects the known sequence-level similarity between promoters and enhancers (Andersson and Sandelin, 2020; Paramo *et al.*, 2026). Applying gkmexplain on the same test set generated per-class per-base contribution scores, which were analysed using TF-MoDISco. For the enhancer class, this recovered a FOS::JUN motif matching JASPAR MA0099.4 (Fig. 1F) (Ovek Baydar et al., 2026; Matrix profile: FOS::JUN - MA0099.4 - JASPAR, 2026). In addition, pertinent motifs were identified for both CTCF and promoter sequences (Fig. S1-S3). Which indicates that mLS-GKM has learnt biologically relevant regulatory sequence features. The same workflow recovered all three planted motifs from a synthetic 3-class benchmark (Supplementary Methods S3; Fig. S2). Together this demonstrates that the multi-class extension, calibrated probabilities, and streamlined gkmexplain that are part of mLS-GKM combine to make a previously fragmented analysis runnable as a single end-to-end pipeline.

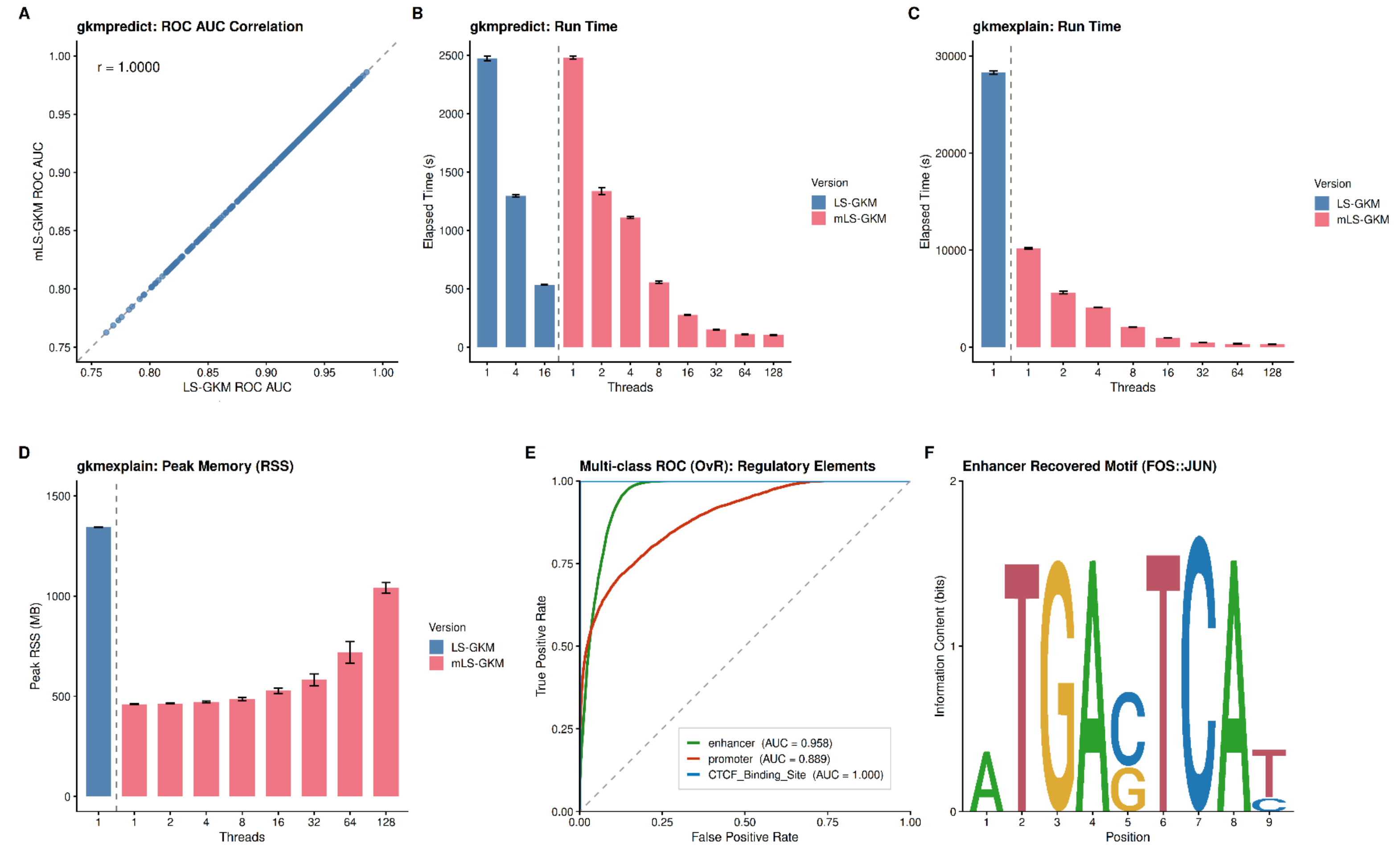

*Figure 1 Performance comparison between LS-GKM and mLS-GKM. A) Comparison of ROC AUC values for 322 ENCODE ChIP-Seq datasets run with LS-GKM and mLS-GKM, showing no change in predictive performance. B) Runtime comparison for gkmpredict, contrasting the original LS-GKM kernel level multithreading with the per-sequence multithreading of mLS-GKM C) Runtime comparison for gkmexplain, showing a 2.8x speed up when both methods are using one thread, due to the more efficient score computation method, along with the increased performance from per sequence multithreading. D) Peak memory usage of gkmexplain, showing the reduction in memory from the more efficient score computations, along with minimal increase in memory with multithreading. E) an example of running mLS-GKM on a multiclass problem, achieving AUCs ranging from 0.889 to 1.000. F) An example motif recovered from running gkmexplain on a multiclass problem.*

## Discussion

mLS-GKM is intended as a drop-in replacement for LS-GKM, but with several practically important extensions. The main contribution is support for multi-class classification with calibrated probabilities, allowing the classification of multiple types of regulatory element while providing interpretable confidence scores. The probability-calibrated mode is the first deployment of pairwise Platt scaling with Wu-Lin-Weng coupling on top of gkm kernels, enabling gkm-SVMs to be used in workflows that require well-defined probability scores. The streaming implemented in gkmexplain reduces the memory requirements and improves runtime without any change to the underlying algorithm. Optional checkpointing makes calibrated multi-class training practical on shared HPC resources by allowing long-running training jobs to survive node failures and wall-clock limits. Together, these changes preserve the strengths of LS-GKM, namely interpretable kernel-based modelling for detecting short regulatory motifs, while removing previous limitations on dataset size and label cardinality.

## Acknowledgements

The analysis was performed on the Cardiff School of Biosciences' Biocomputing Hub HPC infrastructure with resources funded by the Cardiff School of Biosciences.

# Supplementary Material

*mLS-GKM: Efficient Multi-class Regulatory Sequence Classification with Gapped k-mer SVMs*

### Supplementary Methods S1. Datasets and benchmarking protocol

**ENCODE ChIP-seq.** All 333 uniformly processed ENCODE ChIP-seq peak files (The ENCODE Project Consortium, 2012) (source URL) with ≥ 5,000 peaks were downloaded as bigBed and converted to FASTA. For each dataset, an equal number of negative regions was sampled from the genome and rejection-matched to the positives on GC content and repeat fraction. Chromosomes 1 and 2 were held out as a test set, with the remainder used for training. Regions > 1 kb were removed for direct comparability with the original LS-GKM results (Lee, 2016).

**Training hyperparameters.** All models were trained with -t 3 (gkmrbf kernel), -c 10 -g 2 (the recommended RBF settings), -m 4000 (4 GB kernel cache) with 4 threads used to parallelise the kernel computations. Probability calibration (-P) was enabled where indicated. All other parameters were left as the defaults.

**Equivalence test.** Models trained by mLS-GKM and LS-GKM on each ENCODE dataset were compared by diff-ing the model files; the only systematic difference was the internal coding of class labels (1/2 vs 1/−1). Test-set ROC AUCs were computed independently for each implementation.

**Speed and memory benchmarks.** Speed and memory benchmarks were run on a single fixed HPC node (2x AMD EPYC 7543) with all other jobs disabled, using psrecord (Robitaille, 2026) for memory and wall-clock time. For each thread count, experiments were repeated 5 times and the medians reported. Shaded bands in Fig. 1B-D denote interquartile range.

### Supplementary Methods S2. Ensembl Regulatory Build multi-class dataset

The 3-class regulatory-element classification task used annotations from the Ensembl Regulatory Build, release v115 (Zerbino *et al.*, 2016), which integrates ChIP-seq, DNase-seq and other functional-genomics evidence across multiple cell types into a single set of genome-wide regulatory feature calls. Three feature classes were selected:

- **Enhancer regions** (246,403 records), defined in the Regulatory Build by overlap of distal H3K4me1, H3K27ac and open chromatin DNase/ATAC-seq peaks away from annotated transcription start sites.
- **Promoter regions** (35,983 records), open chromatin DNase/ATAC-seq peaks within 100bp of an annotated transcription start site.
- **CTCF Binding Site regions** (90,891 records), defined by direct CTCF ChIP-seq evidence in at least one cell type.

Each class FASTA was constructed by extracting the reference sequence, GRCh38, for each Regulatory Build feature. Chromosomes 1 and 2 were held out as the test set, matching the chromosome-disjoint convention used for the ENCODE benchmark. The total dataset comprises 306,660 training and 117,446 testing sequences across the three classes.

The biological motivation for these three classes is that they exemplify three distinct, well-characterised modes of regulatory sequence: short, motif-dense, position-specific signals at TSS-proximal promoters; longer, more heterogeneous distal enhancers whose sequence determinants overlap partially with promoters (Paramo *et al.*, 2026) and the highly motif-defined CTCF binding sites, which carry a single dominant 19-bp consensus. The marked AUC gap between CTCF (1.000) and Promoter (0.889) in Fig. 1E reflects this gradient of sequence specificity directly. A CTCF Binding Site Motif was recovered use the svmw_emalign script, using the top 1% of probability scores (Fig. S1). Within the set of motifs identified as being important for predicting promoters, both a DCE motif (Fig. S2) and a motif resembling a CAAT box (Fig. S3) were recovered (Lee *et al.*, 2005).

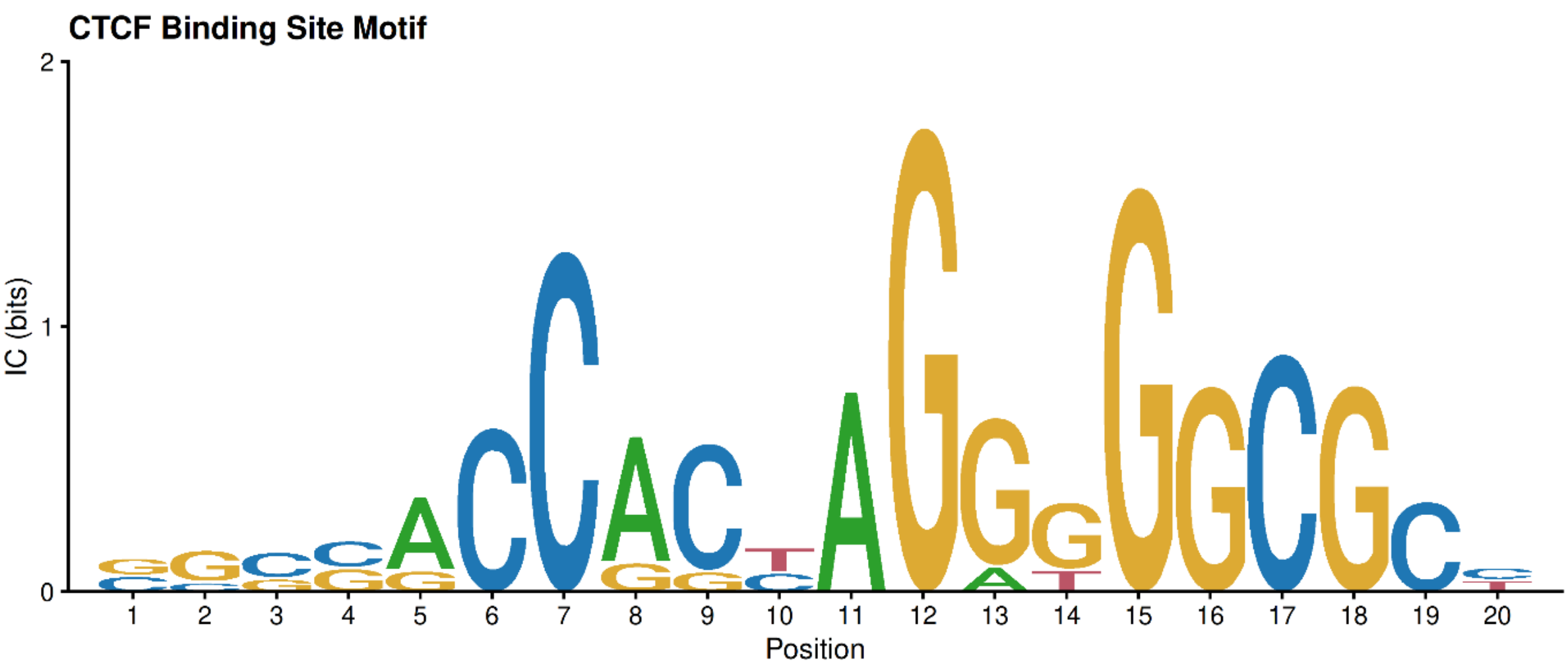

*Figure S1 A recovered CTCF Binding Site Motif, recovered using svmw_emalign script with the top 1% of probability scores*

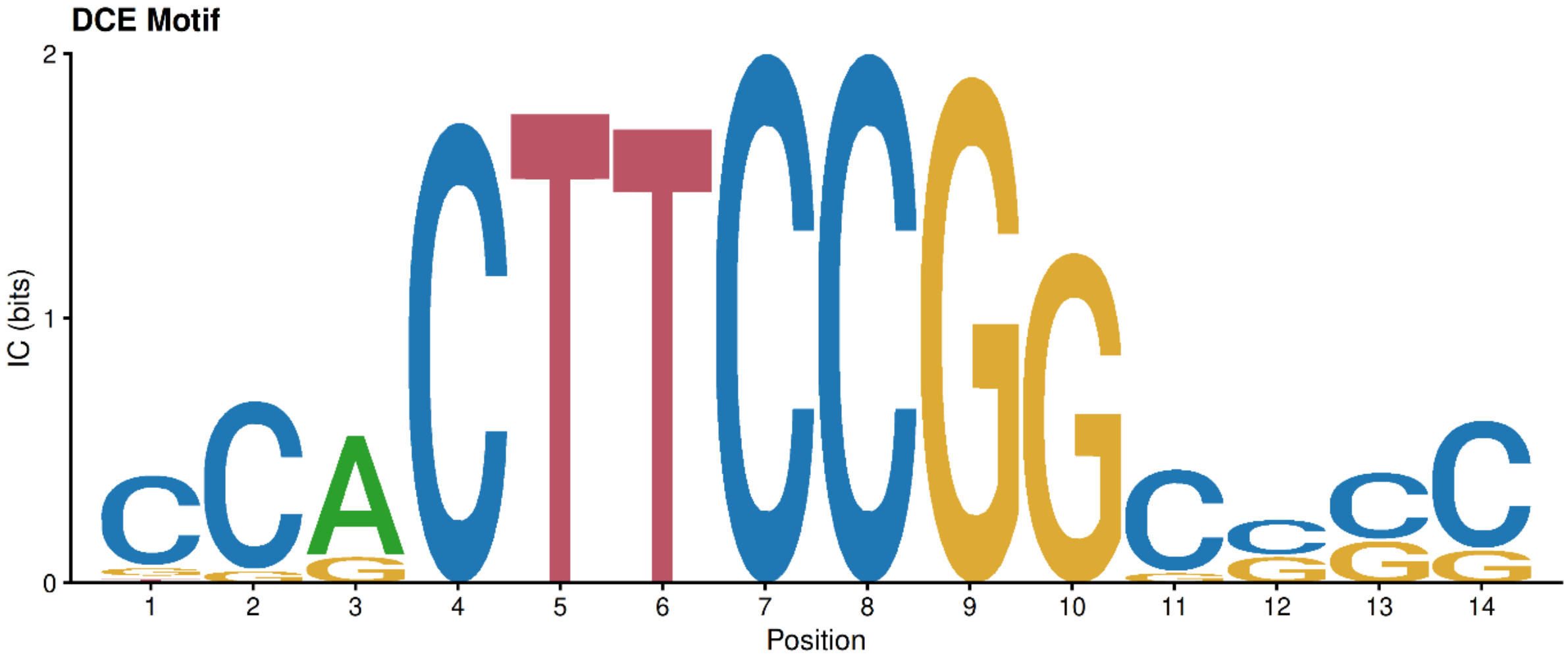

*Figure S2 A recovered Downstream Core Element (DCE) motif (a known core promoter element), recovered using gkmexplain*

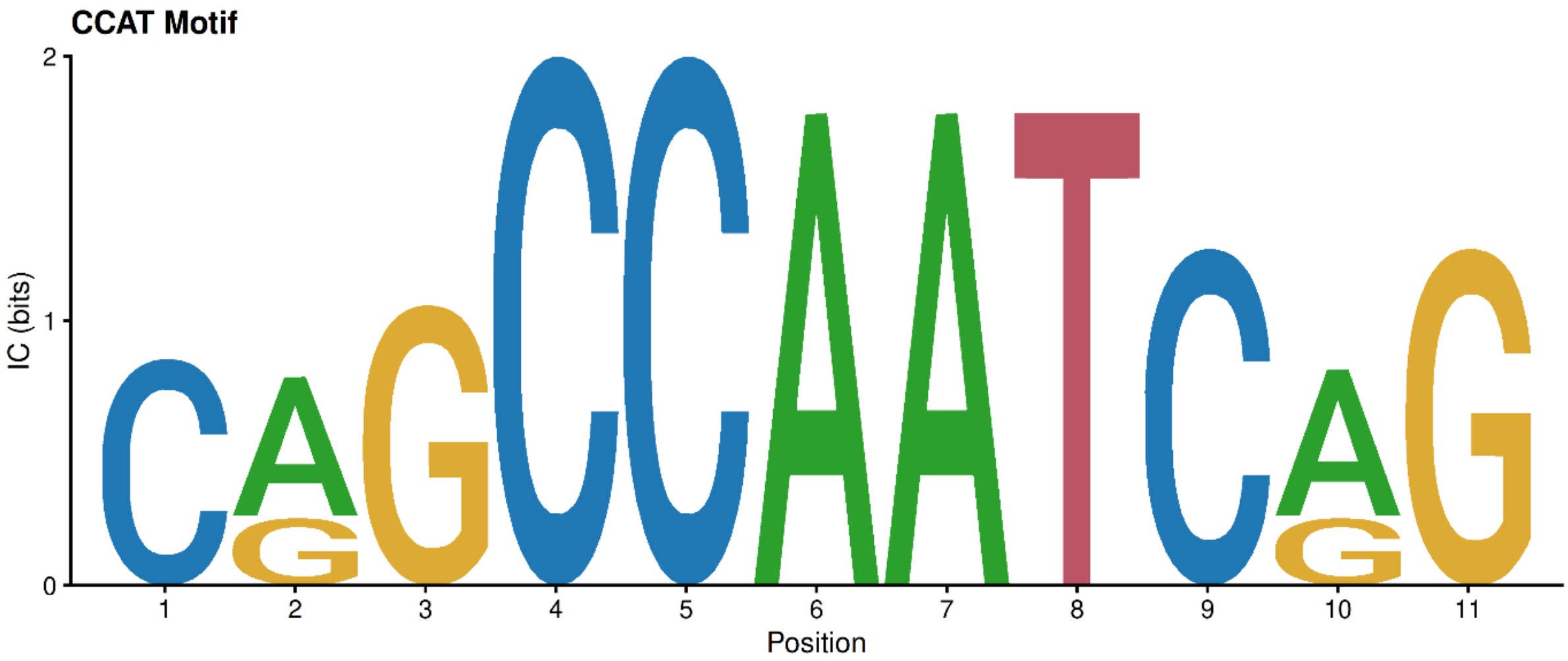

*Figure S3 A recovered CAAT box, recovered using gkmexplain*

**Supplementary Methods S3. Motif Recovery in Synthetic Data**

Three synthetic FASTA files were generated, termed A, B and C. In each file 15,000 sequences of length 500 were generated with background frequencies (A/C/G/T 0.3/0.2/0.2/0.3). Into each record 1-3, non-overlapping, instances of simplified motifs were inserted. Motif A was GTAAACA, a simplified FOXK2 motif (Ovek Baydar *et al.*, 2026; *Matrix profile: FOXK2 - MA1103.2 - JASPAR*, 2026). Motif B was GGGGAGGGG, a simplified SP1 motif (Ovek Baydar *et al.*, 2026; *Matrix profile: SP1 - MA0079.5 - JASPAR*, 2026) and Motif C was CTTATCG, a simplified GATA2 motif (Ovek Baydar *et al.*, 2026; *Matrix profile: GATA2 - MA0036.2 - JASPAR*, 2026).

The FASTA files were then split 70/30 into training and testing files. The training files were used to train a model with -t 3 (gkmrbf kernel), -c 10 -g 2 (the recommended RBF settings) and -P. The testing files were combined into one FASTA and evaluated with gkmexplain, in both -P and -D modes.

In order to recover motifs from the gkmexplain scores, the scores need to be processed through TF-MoDISco, as described in the gkmexplain publication (Shrikumar, Prakash and Kundaje, 2019). Briefly, dinucleotide preserved shuffled sequences were generated (with the make_dnshuff_fasta.py script) and scored with gkmexplain. A set of hypothetical contribution scores were also generated by running the combined testing FASTA file through gkmexplain with the -m 1 flag set. These were then input into TF-MoDISco via a custom script (explain_script.py), which is based on the original gkmexplain interpretation notebooks, but with additional steps to deal with the multiclass output format of mLS-GKM gkmexplain. All three motifs were able to be recovered using the default parameters (Fig. S4).

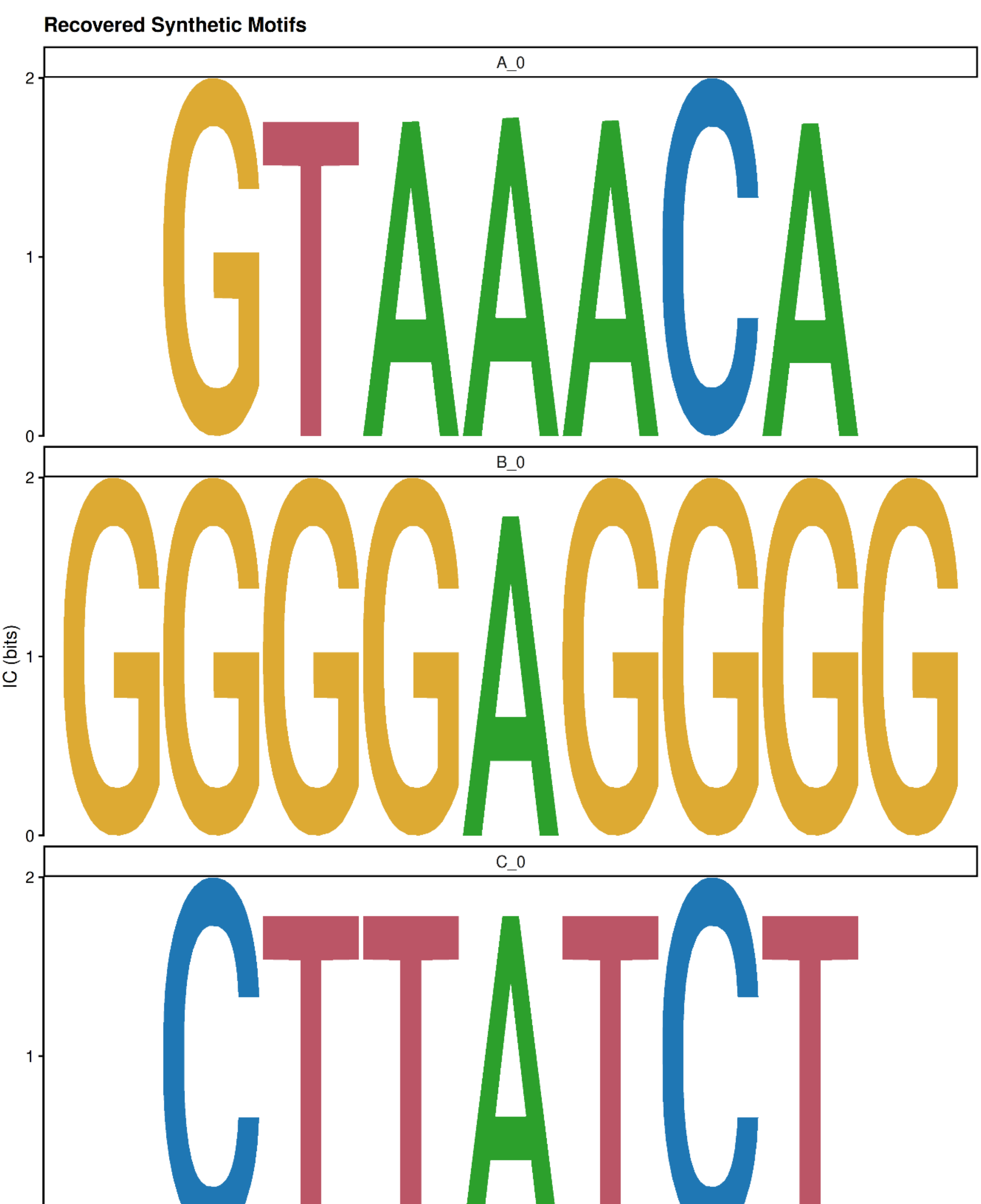

*Figure S4 The three inserted motifs were successfully recovered using gkmexplain and TF-MoDISco.*

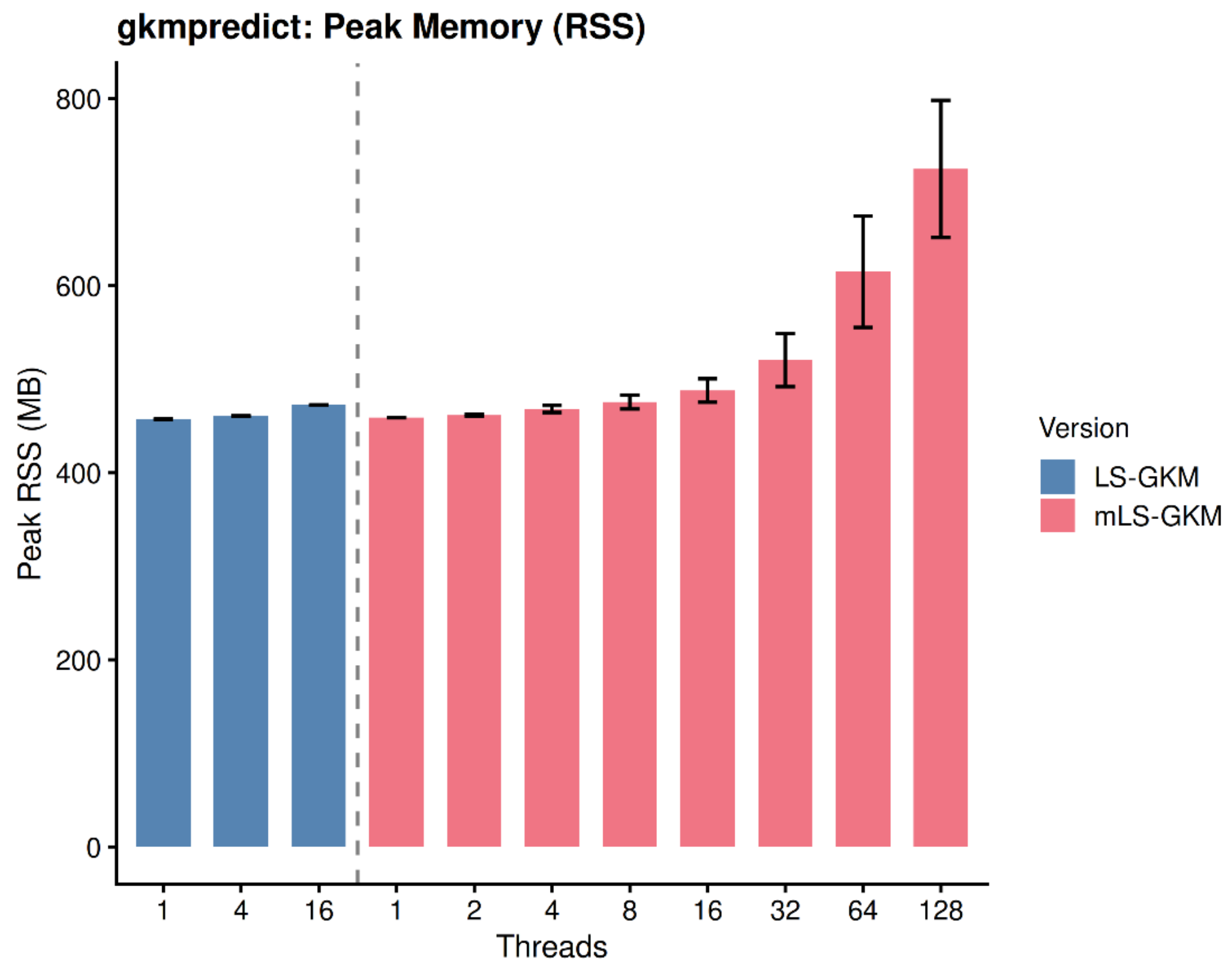

*Figure S5 mLS-GKM gkmpredict memory scaling, showing a moderate increase in memory consumption at high numbers of threads, due to the overhead of managing multiple threads.*